\documentstyle[aps,epsfig,multicol]{revtex}

\begin{document}
\draft
\preprint{\begin{tabular}{l}
\hbox to\hsize{December, 1998 \hfill KAIST-TH 98/20}\\[-3mm]
\hbox to\hsize{hep-ph/9812229 \hfill SNUTP 98-139}\\[5mm] \end{tabular} }


\title{ Probing SUSY-induced CP violations \\  at B factories
}
\author{Seungwon Baek and Pyungwon Ko }
\address{Department of Physics, KAIST \\ Taejon 305-701, Korea}
\maketitle
\begin{abstract}
In the minimal supersymmetric standard model (MSSM), the $\mu-$parameter 
and the trilinear coupling $A_t$ may be generically complex and can affect 
various observables at B factories.
Imposing the edm constraints, we find that 
there is no new large phase shift in the $B^0 - \overline{B^0}$ mixing, 
CP violating dilepton asymmetry is smaller than $0.1 \%$, and   
the direct CP violation in $B\rightarrow X_s \gamma$ can be as large as 
$\sim \pm 16 \%$.
\end{abstract}


\begin{multicols}{2}
\narrowtext

In the minimal supersymmetric standard model (MSSM), there can be many new
CP violating (CPV) phases beyond the KM phase in the standard model (SM). 
These SUSY CPV phases are constrained by electron/neutron electric dipole 
moment (EDM) and have been considered very small ($\delta \lesssim 
10^{-2}$ for $M_{\rm SUSY} \sim O(100)$ GeV ) \cite{susycp}. 
However there is a logical possibility that various contributions to 
electron/neutron EDM  cancel with each other in substantial part of the 
MSSM parameter space even if SUSY CPV phases are  $\sim O(1)$ 
\cite{nath} \cite{kane}. 
Or one can consider effective  SUSY models 
where decouplings of the 1st/2nd generation sfermions are invoked to solve 
the SUSY FCNC/CP problems. Then one loop edm constraints are 
automatically evaded in this scenario \cite{kaplan}.
In such cases, these new SUSY phases may affect $B$ and $K$ physics 
in various manners. 
Closely related with this is the  electroweak baryogenesis (EWBGEN) 
scenario in the MSSM.
One of the fundamental problems in particle physics is to understand the
baryon number asymmetry, $n_B/s = 4 \times 10^{-12}$, and currently popular 
scenario is EWBGEN in the MSSM \cite{carena}.
The EWBGEN is in fact possible in a certain region of the MSSM parameter 
space, especially for light stop ($120~{\rm GeV} \lesssim m_{\tilde{t}_1} 
\lesssim 175$ GeV,  dominantly $\tilde{t}_1 \simeq \tilde{t}_R$) 
with CP violating phases in $\mu$ and $A_t$ parameters.
Then one would expect this light stop and new CP violating phases may  
lead to observable consequences to $B$ physics.

The purpose this letter is to consider a possibility of observing effects 
of these new flavor conserving and CPV phases ($\phi_\mu$ and $\phi_{A_t}$) 
at $B$ factories in the MSSM (including EWBGEN scenario therein).
More specifically, we consider the following observables : 
SUSY contributions to the $B^0 - \overline{B^0}$ mixing, the dilepton CP 
asymmetry in the $B^0 \overline{B^0}$ decays, and the direct CP asymmetry 
in $B \rightarrow X_s \gamma$. The $B^0 - \overline{B^0}$ mixing is important 
for determination of three angles of the unitarity triangle. Also, 
last two observables are vanishingly small in the standard model (SM), 
and any appreciable amounts of these asymmetries would herald the existence 
of new CP violating phases beyond the KM phase in the SM. The question to be 
addressed in this letter is how much these observables can be deviated from
their SM values when $\mu$ and $A_t$ parameters in the MSSM have new CPV 
phases.

In order to study $B$ physics in the MSSM, we make the following assumptions. 
\cite{misiak}. First of all, the 1st and the 2nd family squarks are assumed 
to be degenerate and very heavy in order to solve the SUSY FCNC/CP problems
\cite{kaplan}.  
Only the third family squarks can be light enough to affect
$B\rightarrow X_s \gamma$ and $B^0 - \overline{B^0}$ mixing.
We also ignore possible flavor changing squark mass matrix elements
that could generate gluino-mediated flavor changing neutral current (FCNC)
process in addition to those effects we consider below. Recently, such 
effects were studied in the $B^0 - \overline{B^0}$ mixing \cite{cohen-B} 
\cite{randall}, the branching ratio of $B\rightarrow X_s \gamma$ 
\cite{cohen-B} and CP violations therein \cite{hou} \cite{kkl}, and 
$B\rightarrow X_s l^+ l^-$ \cite{kkl}, respectively. 
Ignoring such contributions, the only source of the FCNC in our model is the 
CKM matrix, whereas there are new CPV phases coming from the phases of $\mu$ 
and $A_t$ parameters in the flavor preserving sector in addition to the KM 
phase $\delta_{KM}$ in the flavor changing sector. In this sense, this paper
is complementary to the ealier works \cite{cohen-B}-\cite{kkl}.

Even if the 1st/2nd generation squarks are very heavy and degenerate, there
is another important edm constraints considered by Chang, Keung and Pilaftsis 
(CKP)  for large $\tan\beta$ \cite{pilaftsis}.  
This constraint comes from the two loop diagrams involving stop/sbottom 
loops, and is independent of the masses of the 1st/2nd generation squarks.
\begin{equation}
( {d_f \over e } )_{\rm CKP} = Q_f {3 \alpha_{\rm em} \over 64 \pi^2}
{R_f  m_f \over M_A^2} \sum_{q=t,b} \xi_q Q_q^2 
F \left( {M_{\tilde{q}_1}^2 \over M_A^2}, {M_{\tilde{q}_2}^2 \over 
M_A^2 } \right)  
\end{equation} 
where $R_f = \cot\beta ~ (\tan\beta)$ for $I_{3 f} = 1/2 ~(-1/2)$, and 
\begin{equation}
\xi_t = {\sin 2\theta_{\tilde{t}} m_t {\rm Im} (\mu e^{i \delta_t} ) 
\over \sin^2 \beta ~v^2}, ~~~~
\xi_b = {\sin 2\theta_{\tilde{b}} m_b {\rm Im} ( A_b e^{-i \delta_b} ) 
\over \sin \beta ~\cos\beta~v^2},
\end{equation}
with $\delta_q = {\rm Arg} (A_q + R_q \mu^* )$, and $F(x,y)$ 
is a two-loop function given in Ref.~\cite{pilaftsis}.
Therefore, this CKP edm constraints can not be simply evaded by making the 
1st/2nd generation squarks very heavy, and it turns out that this puts a 
very strong constraint on the possible new phase shift in the 
$B^0 - \overline{B^0}$ mixing.  

In the MSSM, the chargino mass matrix is given by
\begin{equation}
M_{\chi^{\pm}} = \left( 
\begin{array}{cc}
M_2 & \sqrt{2} m_W \sin\beta
\\
\sqrt{2} m_W \cos\beta  & \mu 
\end{array}
\right) .
\end{equation}
In principle, both $M_2$ and $\mu$ may be complex, but one can perform a phase 
redefinition in order to render the $M_2$ is real \cite{kane}. 
In such a basis, there appears one new phase ${\rm Arg} (\mu)$ 
as a new source of CPV. The stop mass matrix is given by 
\begin{equation}
M_{\tilde{t}}^2 = \left( 
\begin{array}{cc}
m_Q^2 + m_t^2 + D_L   & m_t ( A_t^* - \mu / \tan \beta ) 
\\
m_t ( A_t - \mu^* / \tan \beta ) & m_U^2 + m_t^2 + D_R
\end{array}
\right) ,
\end{equation}
where $D_{L} = ( {1\over 2} - {2\over 3}~ \sin^2 \theta_W ) \cos 2\beta ~
m_Z^2$ and $D_{R} =  {2\over 3} \sin^2 \theta_W \cos 2\beta ~m_Z^2$. 
There are two new phases in this matrix, ${\rm Arg} (\mu)$ and ${\rm Arg} 
(A_t)$ in the basis where $M_2$ is real. 

We scan over the MSSM parameter space as indicated below 
(including that relevant to the EWBGEN scenario in the MSSM) :
$  
80~{\rm GeV} < | \mu | < 1~{\rm TeV},~ 
80~{\rm GeV} < M_2 < 1~{\rm TeV},~
60~{\rm GeV} < M_A < 1~{\rm TeV},~
2 < \tan\beta < 70,~ 
(130~{\rm GeV})^2 <  M_Q^2  <  ( 1 ~{\rm TeV} )^2,~
  - ( 80~{\rm GeV})^2  <  M_U^2  <  (500~{\rm GeV})^2,~
0 < \phi_{\mu}, \phi_{A_t} < 2 \pi,~  0 < | A_t | < 1.5 
~{\rm TeV}.
$
We have imposed the following experimental constraints : 
$M_{\tilde{t}_1} > 80$ GeV independent of the mixing angle 
$\theta_{\tilde{t}}$, $M_{\tilde{\chi^{\pm}}} > 83$ GeV, 
${\rm Br} (B\rightarrow X_{sg}) < 6.8 \%$ \cite{bsg}, and
$0.77 \leq R_{\gamma} \leq 1.15$ \cite{alexander}, 
where $R_{\gamma}$ is defined as 
$R_{\gamma}= BR(B \to X_s \gamma)^{expt}/ BR(B \to X_s \gamma)^{SM}$ and 
$BR(B \to X_s \gamma)^{SM} = (3.29 \pm 0.44) \times 10^{-4}$. 
It has to be emphasized that this parameter space is larger than that in  
the constrained MSSM (CMSSM) where the universality of soft terms at the GUT 
scale is assumed.  Especially, we will allow $m_U^2$ to be negative as well 
as positive,  which is preferred in the EWBGEN scenario
\cite{carena}.  
Since we do not impose any further requirement on the soft 
terms (such as radiative electroweak symmetry breaking, absence of color 
charge breaking minima, etc.), our results of the maximal deviations of 
$B^0-\overline{B^0}$ mixing and $A_{\rm CP}^{b\rightarrow s\gamma}$
from the SM predictions are conservative upper bounds within the MSSM. 
If more theoretical conditions are imposed, the maximal deviations will 
be smaller. 
In the numerical analysis, we used the following numbers for the input 
parameters : $\overline{m_c}(m_c(pole)) = 1.25$ GeV, 
$\overline{m_b}(m_b(pole)) = 4.3$ GeV, $\overline{m_t}(m_t(pole)) = 165$ 
GeV (these are running masses in the $\overline{MS}$ scheme), 
and $| V_{cb} | = 0.0410, |V_{tb}| = 1, | V_{ts} | = 0.0400$ 
and $\delta_{KM} = \gamma (\phi_3) = 90^{\circ}$ for the CKM matrix elements.

The $B^0 - \overline{B^0}$ mixing is generated by the box diagrams 
with $u_i-W^{\pm} (H^{\pm})$ and $\tilde{u}_i-\chi^{\pm}$ running around 
the loops in addition to the SM contribution. The resulting effective 
Hamiltonian is given by 
\begin{equation}
H_{\rm eff}^{\Delta B=2} = - {G_F^2 M_W^2 \over (2\pi)^2}~
\sum_{i=1}^3 C_i O_i,
\end{equation}
where $O_1 = \overline{d}_L^{\alpha} \gamma_{\mu} b_{L}^{\alpha}~
\overline{d}_L^{\beta} \gamma^{\mu} b_{L}^{\beta},
O_2 = \overline{d}_{L}^{\alpha} b_{R}^{\alpha} \overline{d}_L^{\beta} 
b_{R}^{\beta}$, and  
$O_3 = \overline{d}_{L}^{\alpha} b_{R}^{\beta} \overline{d}_L^{\beta} 
b_{R}^{\alpha}$. 
The Wilson coefficients $C_i$'s at the electroweak scale ($\mu_0 \sim M_W 
\sim M_{\tilde{t}}$) can be written  schematically as \cite{branco}
\begin{eqnarray}
C_1 ( \mu_0 ) & = & \left( V_{td}^* V_{tb} \right)^2 ~
\left[ F_V^W (3;3) + F_V^H (3;3) +  A_V^C \right]
\nonumber  \\
C_2 ( \mu_0 ) & = & \left( V_{td}^* V_{tb} \right)^2 ~F_S^H (3;3)
\nonumber  \\
C_3 ( \mu_0 ) & = & \left( V_{td}^* V_{tb} \right)^2 ~A_S^C,
\end{eqnarray}  
where the superscripts $W,H,C$ denote the $W^{\pm}, H^{\pm}$ and chargino
contributions respectively, and  
\begin{eqnarray}
A_V^C  & = & \sum_{i,j,k,l}^{1,2} 
{1\over 4}~G_{(3,k)}^{i} G_{(3,k)}^{j*} G_{(3,l)}^{i*} G_{(3,l)}^{j} 
Y_1 (r_k, r_l, s_i, s_j ),
\nonumber  
\\
A_S^C & = & \sum_{i,j,k,l}^{1,2} 
H_{(3,k)}^{i} G_{(3,k)}^{j*} G_{(3,l)}^{i*} H_{(3,l)}^{j} 
Y_2 (r_k, r_l, s_i, s_j ),
\nonumber  
\end{eqnarray}
Here $G_{(3,k)}^{i}$ and $H_{(3,k)}^{i}$ are the couplings of $k-$th stop 
and $i-$th chargino with left-handed and right-handed quarks, respectively :
\begin{eqnarray}
G_{(3,k)}^{i} & = & \sqrt{2} C_{R 1i}^* S_{t k1} - 
{ C_{R 2i}^* S_{t k2} \over \sin\beta } ~{m_t \over M_W},
\nonumber \\
H_{(3,k)}^{i} & = & {C_{L 2 i}^* S_{tk1} \over \cos\beta } ~{m_b \over M_W},
\end{eqnarray}
and $C_{L,R}$ and $S_t$ are unitary matrices that diagonalize the chargino
and stop mass matrices. Explicit forms for functions $Y_{1,2}$ and $F$'s can 
be found in Ref.~\cite{branco}, and $r_k = M_{\tilde{t}_k}^2 / M_W^2$ and 
$s_i = M_{\tilde{\chi_i^{\pm}}} / M_W^2$. 
It should be noted that $C_2 ( \mu_0 )$ was misidentified as 
$C_3^H ( \mu_0 )$ in Ref.~\cite{demir}.
The gluino and neutralino contributions are negligible in our model.
The Wilson coefficients at the $m_b$ scale are obtained by renomalization
group running. The relevant formulae with NLO QCD corrections at $\mu = 2$ 
GeV are given in Ref.~\cite{contino}.

In our model, $C_1 ( \mu_0 )$ and $C_2 ( \mu_0 )$ are real  relative to the 
SM contribution. On the other hand, the chargino exchange contributions to 
$C_3 (\mu_0 )$ (namely $A_S^C $)are generically complex relative to the 
SM contributions, and  can generate a new phase shift in the 
$B^0 - \overline{B^0}$ mixing relative to the SM value. This effect can be 
in fact significant for large $\tan\beta (\simeq 1/\cos\beta)$, since 
$C_3 (\mu_0)$  is proportional to $ (m_{b} / M_W \cos\beta )^2$ \cite{demir}.
However, the CKP edm constraint puts a strong constraint for large 
$\tan\beta$ case, which was not properly included in  Ref.~\cite{demir}.  
In Fig.~1 (a), we plot $ 2 \theta_d \equiv {\rm Arg}~(M_{12}^{\rm FULL} / 
M_{12}^{\rm SM} )$ as a function of $\tan\beta$. The open squares (the 
crosses) denote those which (don't) satisfy the CKP edm constraints. 
It is clear that the CKP edm constraint on $2 \theta_d$ is in fact very 
important  for large $\tan\beta$, and we have 
$| 2 \theta_d | \lesssim 1^{\circ}$. If we ignored the CKP edm constraint at 
all, then $|2 \theta_d |$ could be as large as $ \sim 4^{\circ}$.  
This observation is important for the CKM phenomenology, 
since time-dependent CP asymmetries in neutral $B$ decays into 
$J/\psi K_S, \pi\pi$ etc. would still measure directly three angles of the 
unitarity triangle even in the presence of new CP violating phases, 
$\phi_{A_t}$ and $\phi_{\mu}$.  Our result is at variance with that obtained
in Ref.~\cite{demir} where CKP edm constraint was not properly included.

If we parametrize the relative ratio of $M_{\rm SM}$ and $M_{\rm SUSY}$ as
$M_{\rm SUSY} / M_{\rm SM} = h e^{-i \theta}$, the dilepton asymmetry is 
given by 
\begin{equation}
A_{ll} = \left( {\Delta \Gamma \over \Delta M } \right)_{\rm SM}
f(h,\theta) \equiv 4~ {\rm Re} (\epsilon_B),
\end{equation}
where $f(h,\theta) = h \sin\theta/(1+2h\cos\theta + h^2)$ and 
$( \Delta \Gamma / \Delta M )_{\rm SM} = (1.3 \pm 0.2 ) \times 10^{-2}$.  
We have neglected the small SM contribution. It is about 
$\sim 10^{-3}$ in the quark level calculation \cite{acuto}, 
but may be as large as $\sim 1 \%$ if the delicate cancellation between the 
$u$ and $c$ quark contribution is not achieved \cite{wolf}.
Since the effects on $\theta_d$ are small in our model, we expect that
the effects on $A_{ll}$ will be similarly negligible \cite{worah}. 
However, we perform a search and confirm that this is indeed the case. 
Scanning over the available MSSM parameter space, we find 
$| f(h,\theta) | \lesssim 0.1$ so that $| A_{ll} | \lesssim 0.1 \%$, which is
well below the current data, $A_{ll} = (0.8 \pm 2.8 \pm 1.2 ) \%$ \cite{opal}.
On the other hand, if any appreciable amount of the dilepton asymmetry is 
observed, it would indicate some new CPV phases in the off-diagonal 
down-squark mass matrix elements \cite{randall}, 
assuming the MSSM is realized in nature. 

On the contrary to the $\theta_d$ and $A_{ll}$ discussed in the previous 
paragraphs, the magnitude of $M_{12}$ is related with the mass difference of 
the mass eigenstates of the neutral $B$ mesons : 
$\Delta m_B = 2 | M_{12}| = (3.05 \pm 0.12) \times 10^{-13}~{\rm GeV}$, 
and thus it will affect the determination of $V_{td}$ from the $B^0 - 
\overline{B^0}$ mixing. We have considered $|M_{12}^{\rm FULL} 
/M_{12}^{\rm SM}|$ and its correlation with 
${\rm Br}(B\rightarrow X_s \gamma)$  are shown in Fig.~1 (b).
The deviation from the SM can be as  large as $\sim 60 \%$, and the 
correlation behaves differently from the minimal supergravity case 
\cite{goto}. 
We repeated the same analyses for $B_s^0 - \overline{B_s^0}$ mixing. There is
no large new phase shift ($2 |\theta_s|$) in this case either, but the modulus 
of $M_{12} (B_s)$ can be enhanced by upto $60\%$ compared to the SM value.  
  
The radiative decay of $B$ mesons, $B\rightarrow X_s \gamma$, is
described by the effective Hamiltonian including (chromo)magnetic dipole
operators. Interference between $b\rightarrow s \gamma$ and $b\rightarrow 
s g$ (where the strong phase is generated by the charm loop via 
$b\rightarrow c\bar{c}s$ vertex) can induce direct CP violation in 
$B\rightarrow X_s \gamma$ \cite{KN}, which is given by 
\begin{eqnarray}
 A_{\rm CP}^{b\rightarrow s\gamma} &  \equiv & 
{ \Gamma ( B \rightarrow X_{\bar{s}} + \gamma ) 
- \Gamma ( \overline{B} \rightarrow X_s + \gamma ) \over
 \Gamma ( B \rightarrow X_{\bar{s}} + \gamma ) 
+ \Gamma ( \overline{B} \rightarrow X_s + \gamma ) }
\nonumber  
\\
& \simeq & {1 \over | C_7 |^2 }~\left\{ 1.23 
{\rm Im}~[C_2 C_7^* ] - 9.52 {\rm Im}
~[C_8 C_7^*] \right.
\nonumber 
\\ & & \left. ~~~~~~~
+ 0.10 {\rm Im}~[C_2 C_8^*] \right\} ~({\rm in }~ \%) , 
\end{eqnarray} 
adopting the notations in Ref.~\cite{KN}.
We have ignored the small contribution from the SM, and assumed that the 
minimal photon energy cut is given by $E_{\gamma} \geq m_B (1-\delta)/2$ 
($\approx 1.8$ GeV with $\delta = 0.3$). $A_{\rm CP}^{b\rightarrow s \gamma}$
is not sensitive to possible long distance contributions
and constitute a sensitive probe of new physics that appears in the short 
distance Wilson coefficients $C_{7,8}$ \cite{KN}. 

The Wilson coefficients $C_{7,8}$ in the MSSM have been calculated by many
groups \cite{bsgamma}, including the PQCD corrections in certain MSSM 
parameter space \cite{giudice}. In this letter, we use the leading order 
expressions for $C_i$'s which is sufficient for $A_{\rm CP}^{b\rightarrow 
s\gamma}$.  After scanning over the MSSM parameter space described in 
Eq.~(3), we find that $A_{\rm CP}^{b\rightarrow s \gamma}$ can be as large 
as $\simeq \pm 16\%$ if chargino is light enough, even if we impose the edm 
constraints. Its correlation with $Br (B\rightarrow X_s \gamma)$ and chargino 
mass are shown in Figs.~2 (a) and (b) respectively.  Our results are 
quantitatively different from other recent works \cite{strumia}\cite{keum}, 
mainly due to the different treatments of soft terms.
In the minimal supergravity scenario, this asymmetry is very small, because 
the $A_t$ phase effect is very small in the electroweak scale \cite{keum}.
If the universality assumption is relaxed, one can accomodate larger direct 
asymmetry without conflicting with the edm constraints.

In conclusion, we studied consequences at $B$ factories in the MSSM for 
the scenario where the first two generation sfermions are heavy, and 
there are CP violating phases in $A_t$ and $\mu$ parameters. 
The main results can be summarized as follows. 
There is no appreciable new phase in the $B^0-\overline{B^0}$ mixing
$( | 2 \theta_d | \lesssim 1^{\circ}$), so that time-dependent  
CP asymmetries in neutral $B$ decays (into $J/\psi K_S, \pi\pi$ etc.) 
still measure essentially three angles of the unitarity triangle even if 
there are new complex phases in $\mu$ and $A_t$ parameters.
The size of the $B^0-\overline{B^0}$ mixing can be enhanced up to 
$\sim 60 \%$ compared to the SM contribution, which will affect 
determination of $V_{td}$ from  $\Delta m_{B}$. 
There is no large shift in ${\rm Re} ( \epsilon_B )$, and 
dilepton CP asymmetry is rather small ($| A_{ll} | \lesssim 0.1 \%$).
Direct CP asymmetry in $B\rightarrow X_s \gamma$ can be as large as 
$\sim \pm 16 \%$ if chargino is light enough. 

These results would set the level of experimental sensitivity that one has to 
achieve in order to probe the SUSY-induced CP violations at $B$ factories 
through $B^0-\overline{B^0}$ and $A_{\rm CP}^{b\rightarrow s\gamma}$ mixing.  
Our results are conservative in a sense that we did not impose any conditions
on the soft SUSY breaking terms except that the resulting mass spectra for
chargino, stop and other sparticles satisfy the current lower bounds from 
LEP and Tevatron. 
Therefore, one would be able to find the effects of the phases of $\mu$ and 
$A_t$ parameters by observing $A_{\rm CP}^{b\rightarrow s\gamma}$ at 
B factories. Other effects of supersymmetric CP violting phases $\phi_{\mu}$
and $\phi_{A_t}$ on $B\rightarrow X_s l^+ l^-$ and $\epsilon_K$ will be
presented in the separate work \cite{baek}.

\acknowledgements
The authors thank A. Ali, G.C. Cho, J. Cline, A. Pilaftsis and O. Vives for 
useful communications. 
This work is supported in part by KOSEF through CTP at Seoul National 
University, by KOSEF Contract No. 971-0201-002-2,  
by Korea Rsearch Foundation Program 1998-015-D00054 (PK), 
and by KOSEF Postdoctoral Fellowship Program (SB).


%
%


%
\begin{figure}
    \begin{center}
      \begin{picture}(140,170)
   \put(-50,0){\epsfig{file=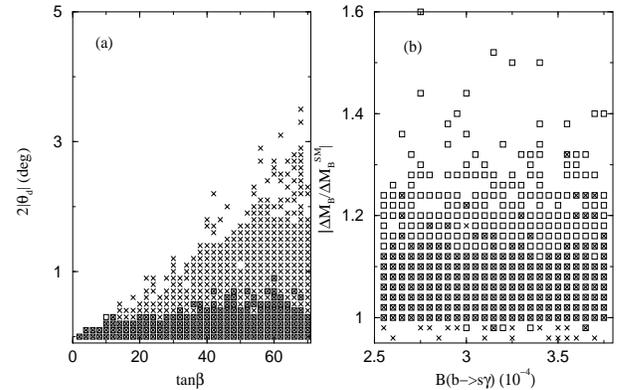,width=8cm,height=6cm}}
        \end{picture}
     \end{center}
\caption{
Correlations between
(a) $\tan\beta$ vs. 2 $|\theta_d | $, and 
(b) ${\rm Br} (B \rightarrow X_s \gamma)$ vs. 
$A_{12}^{\rm FULL} / A_{12}^{\rm SM}$. 
The squares (the crosses) denote those which (do not) satisfy the CKP edm 
constraints.  
}
\label{fig1}
\end{figure}

\begin{figure}
    \begin{center}
      \begin{picture}(140,170)
      \put(-50,0){\epsfig{file=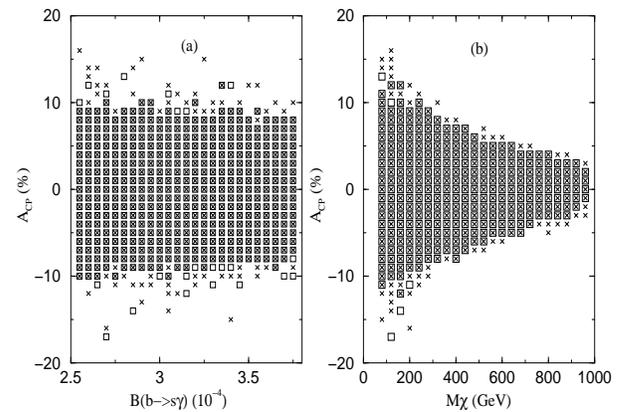,width=8cm,height=5.5cm}}
        \end{picture}
     \end{center}
\caption{
Correlations of $A_{\rm CP}^{b\rightarrow s\gamma}$ with 
(a) ${\rm Br} (B \rightarrow X_s \gamma)$ and (b) the lighter chargino
mass $M_{\chi^{\pm}}$.
The squares (the crosses) denote those which (do not) satisfy the CKP edm 
constraints. 
}
\label{fig2}
\end{figure}

\end{multicols}
\vfil\eject
\end{document}